\documentclass[a4paper]{jpconf}
\usepackage{braket,graphicx,braket,units}
\begin{document}
\title{Quantum computing for data science}

\author{Barry C.\ Sanders}

\address{Institute for Quantum Science and Technology,
	University of Calgary, Alberta T2N~1N4, Canada}

\ead{sandersb@ucalgary.ca}

\begin{abstract}
I provide a perspective on the development of quantum computing for data science, including a dive into state-of-the-art for both hardware and algorithms and the potential for quantum machine learning.
\end{abstract}

\section{Introduction}
Quantum information theory
transforms the very foundations of information theory and computing~\cite{NC10,Aar13}.
The pre-quantum (known as `classical') scientific framework allows objective information
to be labelled by integers such as bit strings
(e.g., representing text characters by 7-bit strings according to the American Standard Code for Information Interchange, or ASCII),
which is foundational to information theory.
Processing information can be executed under the rules of Boolean logic,
manifested as concatenations of one-bit operations such as NOT and two-bit operations such as NAND.

Quantum information changes the information game entirely by allowing coherent superpositions of informational states,
which, following the quantum principle of complementarity,
can be thought of as being both corpuscular (particle-like) and undular (wave-like).
For example, a three-bit string 010 becomes quantumly a label for a quantum state~$\ket{010}$
(element of Hilbert space),
which could be manifested physically, say,
by three electrons that are spin-down, spin-up and spin-down,
with a spin-down state labelled~$\ket0$
and a spin-up state labelled~$\ket1$
in Dirac, or bra-ket, notation~\cite{Dir39}.
Superposing this three-electron state with its orthogonal complement would be $\ket{101}$.
For state normalisation implicit throughout this paper,
the superposition of these two states is
$\ket{010}+\ket{101}$,
which,
in binary representation,
is a superposition of the numbers~2 and 5.

These superpositions of information states can be processed quantumly,
i.e., in a way that preserves coherence.
Ideally,
this superposition state can be transformed by an arbitrary unitary map (isometry on Hilbert space).
Realistically,
open-system effects such as noise and loss can impede performance,
but almost-unitary maps
(such as completely positive trace-preserving maps~\cite{NC10} that are close to being unitary)
can suffice for useful quantum-information processing provided that the quantum version of error correction is employed in a fault-tolerant way~\cite{Got21}.

An early motivation for quantum computing was for simulating physics,
particularly for simulating quantum systems in a way that is natural for quantum descriptions~\cite{Fey82},
i.e., by using quantum computing.
Since that original idea,
remarkable quantum algorithms have arisen,
where remarkable refers to delivering superior performance compared to classical algorithms,
such as efficient computation meaning computational resources such as run-time and number of bits or qubits, being quantum bits, scale no worse than polynomially with respect to the size of the problem quantified by how many bits are required for specifying the problem instance~\cite{NC10}.

One example of quantum algorithms being advantageous is for efficiently solving
number factorization,
whose hardness is subexponential (almost exponential) using the best known classical algorithm~\cite{Sho94}.
Another example is a provably quadratically faster algorithm for  function inversion (hence, unstructured search)~\cite{Gro96}.
Quantum annealing is speculated to enhance  optimisation methods in some cases~\cite{HRIT15}.
Building quantum computers is rapidly advancing,
albeit small-scale and noisy,
making quantum computing now viable (in a limited way)
and commercial~\cite{San17,San20,MSSM20}.

Quantum computing is arguably a disruptive technology,
i.e., a technology that is both innovative and market-changing~\cite{Chr97},
if quantum computing eventually delivers a game-changing application
such as for data science~\cite{Chr97,WWE19}.
Some big `ifs' stand in the way of quantum computing being disruptive, though.
My view is that quantum computing faces two huge `ifs', i.e., challenges:
(1)~building a quantum computer
and (2)~making said quantum computer useful.

In a way, the latter is harder than the former.
Scaling a quantum computer to large size (many qubits) and large depth (able to pass through many cycles of quantum logic gates, defined as unitary or near-unitary maps of quantum information states)
is technically extremely challenging but conceivable.
Figuring out,
on the other hand,
what to do with a quantum computer that is truly useful and transformative is, at best, challenging because of a limit to our imagination of what could be the next great quantum algorithm
and, at worst, hopeless if no new great quantum algorithm exists.
Thus, quantum computing is a wonderful, scary, risky adventure,
which we need to navigate with our eyes wide open.
\section{Nexus between quantum computing and data science}
Now I elaborate on the nexus between quantum computing and data science,
as any meaningful overlap between these two galvanising research fields is exciting.
Data science~\cite{KD19} concerns the full pipeline from capturing and maintaining data to processing and analysing these data and,
finally,
communicating or executing data-driven action~\cite{MLW+19}.
Computing is a key part of this pipeline,
not only for exploiting quantum speedup~\cite{RWJ+14} or superior accuracy~\cite{WWDM20}
but also for the possibility of storing information in superposition in a quantum random access memory~\cite{Ble10}
as well as considering cyberattacks on data storage and processing.

Secure data storage could involve quantum cryptography~\cite{PAB+20},
and quantum-secure use of servers could need to exploit methods such as blind quantum computing~\cite{Fit17} or quantum homomorphic encryption~\cite{BJ15,DSS16}.
The full impact of quantum information processing on data science is not yet known and needs extensive study.
Here, we focus here strictly on the impact of quantum computing per se on data science.

I mentioned a little about quantum computing being about superpositions of information states and processing by unitary or near-unitary maps.
As quantum computing is rather subtle in its nature,
let us now take a moment to appreciate how it works.
I would like to illustrate how quantum computing works by considering a quantum simulation,
executable on a quantum computer,
of Schr\"{o}dinger's famous example of a cat
whose existential state can be in a life-death superposition,
so to speak~\cite{Gri14}.
Our purpose is not to philosophise here but rather to recast Schr\"{o}dinger's cat paradox~\cite{Sch35} as quantum information processing thereby illustrating how quantum computing works.

Here, I cast the cat's quantum state of being alive ($b=0$) or dead ($b=1$) as being represented by a quantum state~$\ket{b}$.
Similarly,
we can consider a radioactive nucleus,
that decays by emitting an alpha ($\alpha$) particle.
$\alpha$ decay can be regarded as the emission of an~$\alpha$ particle,
equivalent to a He$^{2+}$ ion,
by quantum tunnelling out of the remaining nucleus~\cite{Gam28}.
The undecayed nucleus can be regarded as being an electromagnetic potential trap for the~$\alpha$ particle,
with the nuclear interior being attractive and the nuclear boundary being repulsive.
We represent the undecayed nucleus by~$\ket0$ and the decayed nucleus by~$\ket1$.
Left alone the initial nuclear state~$\ket0$
evolves into a superposition such as $\ket\pm:=\ket0\pm\ket1$
over one half-life
with choice of~$\pm$ indicating the coherence phase between not having decayed and having decayed.

\begin{figure}[h]
\begin{minipage}{21pc}
\includegraphics[width=16pc]{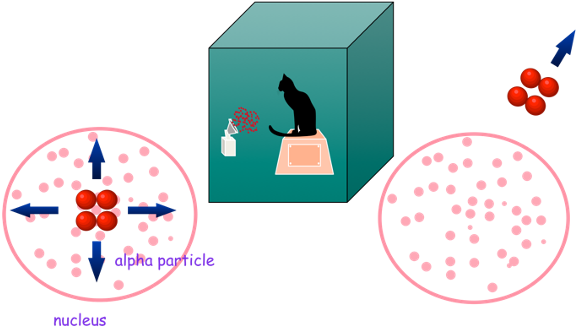}
\caption{A cat is placed in an opaque box along with a poison gas that is released contingent on detecting the decay of an~$\alpha$ particle.
The decay is modelled as the escape of an~$\alpha$ particle,
equivalently a He$^{2+}$ ion,
from the nucleus,
with this decay occurring by quantum tunnelling.}
\end{minipage}\hspace{2pc}%
\begin{minipage}{16pc}
\includegraphics[width=14pc]{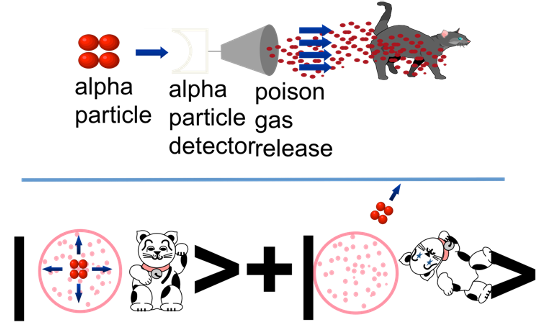}
\caption{Above:
the~$\alpha$ particle is deteced, which causes the release of a poison gas that kills the cat;
in Dirac notation,
the nuclear-cat entangled state is a superposition of the nucleus being undecayed and the cat being alive with the orthogonal state of the nucleus having decayed and the cat having died.}
\end{minipage}
\label{fig:cat}
\end{figure}

To make clear the connection with quantum information processing,
Schr\"{o}dinger's cat paradox is depicted in Fig.~\ref{fig:cat}
as a cat in a box
with poison gas triggered by an $\alpha$-particle detection.
The correlation between~$\alpha$ decay and the cat dying is shown in Fig.~2,
with the resultant entangled nuclear-cat state shown in Fig.~2 as well.
We can write this entangled state as $\ket{00}+\ket{11}$.
This state is extremely important in quantum information science and is often called a Bell state.
Just as we have resources like qubits and gates,
this resource is sometimes called an ebit, short for a pair of maximally entangled qubits.

We can think of Schr\"{o}dinger's cat paradox in terms of a quantum circuit.
The input is initialised to~$\ket{00}$
for an undecayed nucleus and a living cat.
Then one waits for precisely one half-life to achieve $\ket0\mapsto\ket+$.
If decay were a unitary process for one degree of freedom
(not actually the case for nuclear decay but didactically convenient)
then decay could be described by a quantum logic gate known as the Hadamard gate (H),
whose unitary implies
H$:\ket1\mapsto\ket-$ as well,
which would mean that a decayed nucleus implausibly returns to its undecayed state.

The killing of the cat depicted in Fig.~2
is effecting the quantum logic gate known as CNOT, which stands for the controlled-not operation.
The CNOT gate is equivalent to the XOR gate:
flip the second qubit, i.e., apply the quantum NOT gate, if the first qubit is~$\ket1$ but not if the first qubit is~$\ket0$.
Mathematically,
for any two qubits in the computational basis,
written as $\ket{b\;b'}$ for any bits $b$ and~$b'$, the mapping is
${\rm CNOT}:
	\ket{b\;b'}
		\mapsto
			\ket{b\;b'\oplus b}$.
The CNOT gate can be extended in an obvious way,
as a NOT gate controlled by more than one qubit,
such as the controlled-controlled-NOT, or Toffoli, gate
${\rm CCNOT}:
	\ket{b\;b'\;b''}
		\mapsto
			\ket{b\;b'\;b'b''\oplus b}$,
which is valuable as a quantum version of repetition codes~\cite{Bac06}.

For Schr\"{o}dinger's cat paradox,
waiting a half-life means the first qubit is a superposition of~$\ket0$ and~$\ket1$,
not specifically one of these two states, and the output is then the Bell state.
Bell-state preparation,
described by the procedure to manufacture the Schr\"{o}dinger cat paradox,
is a workhorse of some quantum computers,
such as for ion-trap quantum computing~\cite{MS99}.

The Hadamard gate plays a key role in quantum computing.
An outer product of Hadamard gates maps a qubit string by
$\ket{00\cdots0}\mapsto\ket{++\cdots+}$,
which is a superposition of all strings of qubits of given length.
Some quantum algorithms commence this way,
which is at the heart of some references to quantum computing as parallelisation,
but parallelisation is only part of the quantum computing story~\cite{Aar13}.

\section{Building a quantum computer}
Of course we would not be performing quantum computing using cats
(although this terminology is used popularly for quantum computing~\cite{Gri14},
and the term ``two-component Schr\"{o}dinger cat states'' is used for some type of superconducting quantum computing~\cite{MLA+14}
although I prefer the term ``entangled coherent state'' for such states~\cite{San12})
and instead resort to realisable quantum computers.
I now list the current favourite platforms with examples of how the qubits are manifested~\cite{San17,San20},
and I sidestep the growing and intriguing areas of so-called continuous variable quantum computing~\cite{LB99,BSBN02}
and qudit quantum computing~\cite{WHSK20}.

Photonic quantum computing was an early candidate and remains a strong contender.
A single-rail qubit could be manifested as a superposition of no photon and one photon in a particular spatiotemporal mode;
in contrast a dual-rail qubit is a single photon with the logical qubit inherent in the polarisation or which-path or which-time
($\ket{{\rm early}}$ and $\ket{{\rm late}}$, say)
state of the photon.
In magnetic resonance, the qubit can be the nuclear spin state, which can be controlled through hyperfine interactions with electronic qubits in the atom.
Trapped ions can be superpositions of atomic excited and ground states,
and controlled quantum logic gates can be achieved by exploiting collective motion of the ions leading to phononic sidebands of the atomic energy states.
Neutral atomic qubits are also superpositions of ground and excited electronic states,
with entangling two-qubit gates achievable by controlled collisions between atoms.
Solid-state systems could be semiconductors,
with the qubit inherent in the spin of a trapped electron,
and superconducting quantum computing can involve quantized flux or superpositions of charge or using microwave photons in the resonator.
A hybrid semiconductor-superconductor system could lead to topologically protected qubits.

Quantum computers can be realised according to different approach to processing quantum information.
The gate `model' is about making a universal primitive set of gates, such as~H and~CNOT discussed earlier, as well as another gate called~T, which maps~$\ket0\mapsto\ket0$ and $\ket1\mapsto{\rm e}^{{\rm i}\nicefrac{\pi}4}\ket1$.
Any multi-qubit unitary map can be decomposed efficiently into an efficiently scaling number of computer cycles,
with each cycle comprising these single-qubit gates, including the identity gate, plus CNOT gates.

Measurement-based quantum computing, on the other hand, begins by entangling all qubits together and then carving out a circuit by measuring unwanted qubits in an appropriate basis.
The remaining qubits are sequentially measured,
with measurement basis determined by previous measurement outcomes;
the final measurement outcomes provide the desired solution.
Adiabatic quantum computing is about slow, controlled, continuous-time evolution of a Hamiltonian to realise approximately the target unitary map.
Topological quantum computing involves braiding anyonic worldlines and could be realised via the fractional quantum Hall effect.
This list describes universal quantum computers;
purpose-built quantum computers such as quantum annealing---adiabatic or diabatic~\cite{CL21}---and boson sampling~\cite{AA13} enrich the set of quantum computers in development.

Quantum computer technology is quite impressive today, with quantum annealing reaching thousands of qubits and gate-based quantum computing having over a hundred qubits.
Quantum computing primacy has arguably arrived~\cite{DHKL20,San21}.
Certainly, quantum computing technology has surpassed what I imagined would be the status today,
but much more is needed to make quantum computers genuinely useful.

\section{Applied quantum algorithms for data science}
A plethora of quantum algorithms has been developed~\cite{QAZ}.
Although factorisation is the most stunning of the quantum algorithms,
near-term algorithms are the current focus.
Two especially enticing approaches to exploiting noisy intermediate-scale quantum (NISQ) computers~\cite{Pre18}
involve optimisation~\cite{MBB+18},
which can use variational quantum algorithms~\cite{MBB+18,CAB+21},
and quantum machine learning~\cite{SSP15}.
In both cases, a quantum advantage remains an open question.

A business analysis of the potential for commercially viable quantum computing commercially is in combinatorial optimisation~\cite{BGM21}.
This analysis identifies three business verticals,
which are niches that serve needs of a specific set of customers.
They identify the verticals of financial portfolio management,
computing lowest-energy configurations with applications to material design and to pharmaceuticals,
and predicting rare failures for advanced manufacturing.
The near-term gain would, one hopes, arise from exploiting quantum-inspired algorithms~\cite{ADBL20}.
Intriguingly,
even without a quantum advantage,
a theoretical study suggests that a quantum economic advantage for companies developing quantum computers
can emerge even if quantum computers do not offer a quantum computational advantage~\cite{BGM22}.
This argument for a quantum computational advantage,
even if quantum computers do not pay off,
is based on modeling a duopoly comprising a quantum computing company vs a classical computing company.
Their model shows a ``quantum economic advantage'',
even without quantum computing advantage,
partially through market creation.

Quantum machine learning aims to achieve quantum-enhanced machine learning, such as enhancing unsupervised, supervised, generative or transfer learning.
Here, 
I summarise Wiebe's list of key questions that a ``quantum machine learner'' should ask~\cite{Wie20}.
Wiebe relates quantum-enhanced machine learning specifically to certain potential advantages over using classical computing.
He identifies four sought-after quantum enhancements to machine learning, namely, 
fewer computational steps for training or classifying,
reducing sample complexity,
generating a new data model,
and using quantum optimisation for loss functions.
As a second track of quantum machine learning research,
Wiebe summarises how a quantum device could be natural for classifying or extracting features.
To this end,
he lists examples,
namely,
quantum Boltzmann machines and quantum neural networks,
quantum algorithms such as for principal component analysis, and
a quadratic time reduction for quantum-enhanced function invertion applied to data-training.

Consultants are involved in prognosticating commercial applications of quantum algorithms, and I share a little bit here of such endeavours.
Gartner, Inc, anticipates various quantum computing potential applications~\cite{Gar19}.
They regard chemistry applications as first past the post, 
with such applications requiring somewhere around one hundred to two hundred qubits.
As the number of qubits and quantum computational cycles increase,
they forecast quantum optimisation and then quantum machine learning as being viable,
followed by material science,
with each of these application areas needing hundreds to thousands of qubits.
These forecasts are in line with the business forecast discussed above~\cite{BGM21}.
Gartner's analysis culminates with the expectation that, eventually, quantum computers will sovle ``unknown problems'' requiring hundreds of thousands of qubits.

A November 2020 TechRepublic report summarises six experts' predictions for 2021~\cite{Gre20}
beginning with IBM predicting it will achieve 127 qubits,
which indeed happened in 2021.
The aforementioned Gartner, Inc, foresaw that cloud providers would incorporate quantum capability, which is so.
KnowBe4,
a company that focuses on security awareness training,
predicted that quantum computing would break traditional public-key cryptography;
did not happen, and I do not see this happening in the foreseeable future.
Lux Research foreshadowed advances in optimising quantum hardware to reduce resource needs for running quantum algorithms.
Finally, Forrester predicted that 2021 would be a trough of disillusionment,
which is the third of five phases
of the Gartner hype cycle~\cite{FR08,SL10}.
The hype cycle is a plot of maturity or visibility of an emerging technology as a function of time,
and these five phases are
the ``technology trigger'',
followed by the ``peak of inflated expectations'',
then the ``trough of disillusionment'',
subsequently the ``slope of enlightenment''
(waning interest with failure to deliver
with survival dependent on satisfying early adopters)
and, finally, the ``plateau of productivity''.

\section{Conclusions and outlook}
We have reviewed the idea and advantages of quantum computing, the connection with data science emphasising the computational side such as optimisation and machine learning, and took an excursion into what consulting firms are saying.
Now we take stock of all this information to figure out where quantum computing is today and how it affects data science and its applications and verticals.

Clearly, quantum computers are a game changer for computing, at least at a conceptual level.
The idea that hard problems can be solved efficiently on a quantum computer shows that quantum computing is more powerful,
at least in one type of application.
That quantum computing is \emph{provably} quadratically superior,
by some measure,
at inverting functions,
also points to quantum computing being superior in some way to classical computing.

On the other hand, quantum computers today are small and noisy,
and, even if they were large and coherent,
their application to data science entails only speculative advantages.
Whether or not quantum computers are advantageous for data science might only be resolved by building quantum computers, testing algorithmic performance on such computers,
and seeing whether an advantage has been found or not;
this empirical approach is already ongoing in the field.
Fortunately, quantum-inspired algorithms,
which typically arise as ways to challenge a purported quantum advantage by solving the same problem as well or better on a classical computer,
are giving beneficial computational results at present that arguably would not be happening if quantum computing research were not occurring.

Given the unknowns of quantum computing,
including how much quantum computers can scale up and whether we will have good algorithms that exploit the potential of quantum computing for data science,
an approach I call ``quantum-aware, quantum-ready, quantum-active'' is prudent.
This approach means that, at this stage,
we stakeholders in quantum computing, including data scientists, need to be aware of when and how quantum computing could change the game.
We maintain awareness by testing how well existing quantum computers or simulators thereof perform on problems of interest.
If and when quantum computing matters for the application at hand, 
it is time to become quantum-ready, which means train up on the technology and prepare for transitioning to quantum computing so that, when the time comes, we are ready to use it without delay.
Finally, in the optimistic scenario that quantum computing has become a disruptive technology whose adoption is necessary,
we have reached the quantum-active stage where data scientists need to be fully engaged in using quantum computing to solve problems.
We need to remember that quantum computing is a high-risk high-reward venture and treat it accordingly.

\section*{Acknowledgments}
This project has been funded by the Alberta Government and by NSERC.
Thanks to S.\ L.\ Sanders who prepared the figures over two decades ago;
those figures have been valuable for my presentations on Schr\"{o}dinger's cat ever since but have not been published before.

\section*{References}
\bibliography{qdata}
\end{document}